\documentclass[twocolumn,10pt,amsmath,amssymb]{revtex4}
\usepackage{graphicx}
\usepackage{bm}
\usepackage{epsf}

\begin{document}

\title{FUSION REACTIONS IN MULTICOMPONENT DENSE MATTER}

\author{D.~G.~Yakovlev}
    
\affiliation{
Ioffe Physico-Technical Institute, Poliekhnicheskaya 26, 194021    
Saint-Petersburg, Russia \\
Department of Physics $\&$ The Joint Institute    
for Nuclear Astrophysics, University of Notre Dame,  Notre Dame,    
IN 46556 USA 
}    

\author{L.~R.~Gasques } 

\affiliation{
Department of Physics $\&$ The Joint Institute    
for Nuclear Astrophysics, University of Notre Dame,  Notre Dame,    
IN 46556 USA}

\author{ A.~V.~Afanasjev } 

\affiliation{Department of Physics and Astronomy,
Mississippi State University,
P.O.~Drawer 5167, MS 39762-5167 USA}

\author{ M.~Beard,  M.~Wiescher } 

\affiliation{
Department of Physics $\&$ The Joint Institute    
for Nuclear Astrophysics, University of Notre Dame,  Notre Dame,    
IN 46556 USA}

\begin{abstract}
We analyze thermonuclear and pycnonuclear fusion reactions in dense matter
containing atomic nuclei of different types.
We extend a phenomenological expression for the reaction
rate, proposed recently by Gasques {\it et al}.\ \cite{leandro05}
for the one-component plasma of nuclei, to
the multi-component plasma. The expression contains several fit
parameters which we adjust to reproduce the best microscopic
calculations 
available in the literature. Furthermore, we show that
pycnonuclear burning is drastically affected by an (unknown) structure of
the multi-component matter (a regular lattice, a
uniform mix, etc.). We apply the results to study nuclear
burning in a $^{12}{\rm C}^{16}{\rm O}$ mixture. In this context
we present new calculations of the astrophysical $S$-factors for
carbon-oxygen and oxygen-oxygen fusion reactions. We show
that the presence of a CO lattice
can strongly suppress carbon ignition in white dwarf cores
and neutron star crusts
at densities
$\rho \gtrsim 3 \times 10^9$ g~cm$^{-3}$ and temperatures $T
\lesssim 10^8$~K.
\end{abstract}


\maketitle

\section{Introduction}
\label{introduction}

Nuclear reactions are most important for the physics of
stars. They determine hydrogen burning in main-sequence stars,
helium burning in red giants, and carbon, neon, and oxygen burning
at later stages. They 
determine also nucleosynthesis in shock driven stellar explosions,
such as type II supernovae, as well as ignition and burning
in accreting stars. 

Steady-state and explosive thermonuclear carbon burning 
during late stellar evolution
\cite{eleid-2004} and in shock fronts of type II supernovae
\cite{pardo-couch-arnett-74}
is governed by the $^{12}$C+$^{12}$C and possibly by the
$^{12}$C+$^{16}$O fusion processes.
Similarly, thermonuclear
oxygen burning is mainly determined by the $^{16}$O+$^{16}$O and
possibly by the $^{16}$O+$^{20}$Ne  reactions
\cite{truran-arnett-1970}. The ignition and nucleosynthesis
during these burning phases critically depend on the
initial fuel abundance and on the thermonuclear
reaction rates.

In high-density cores of white dwarfs and crusts of
neutron stars, the thermonuclear reactions are strongly affected by
plasma effects 
(especially important for carbon ignition
in cores of accreting massive white dwarfs 
for triggering
type Ia supernova explosions). The ignition conditions are
sensitive to the $^{12}$C and $^{16}$O abundance 
and to the fusion reaction
rates 
\cite{niemeyer-woosley-1997}.

Carbon ignition has also been suggested as a trigger
of superbursts in surface layers of accreting
neutron stars \cite{strohmeyer-bildsten}. However, the required ignition
conditions seem to disagree with the observed
superburst lightcurves \cite{cumming-2005}. Alternative
explanations are presently being discussed, such as carbon
ignition in the crust of an accreting strange (quark) star, to accommodate
the observed light curve characteristics \cite{page-cumming-2005}.
While in most of the scenarios pure carbon burning
dominates the energy production, ignition conditions and
associated nucleosynthesis are affected by 
the presence of other elements.

Pycnonuclear burning occurs in dense and cold
cores of
white dwarfs \cite{svh69} and in  
crusts of accreting neutron stars \cite{Schramm, hz}.
Theoretical formalism has been mostly developed for pycnonuclear
reactions between equal nuclei, but 
one often needs to consider a multi-component matter,
for instance,
carbon-oxygen cores
of white dwarfs.

In a previous publication
\cite{leandro05} we have focused on fusion
reactions between equal nuclei in a one-component plasma (OCP) of
atomic nuclei (ions). In the present work we expand 
the study towards a multi-component plasma (MCP).
The problem  has two
aspects; the first one is associated with the underlying
nuclear physics while the second one is concerned with the plasma
physics.
The nuclear part deals with the reliable determination of
astrophysical $S$-factors at stellar energies. 
These energies are low (typically, lower than a few MeV), 
in particular if compared to the presently accessible
range of low-energy fusion experiments.
This prevents direct measurements
of $S$-factors at laboratory conditions. Thus, one needs to calculate
the $S$-factors theoretically and use these results
to extrapolate measured $S$-factors towards lower stellar energy range.
In Section \ref{nuclear} we present calculations of
the $S$-factors for the two reactions of astrophysical importance,
$^{12}$C+$^{16}$O and $^{16}$O+$^{16}$O.

The plasma physics problem consists in calculating the Coulomb
barrier penetration in nuclear reactions taken into account
Coulomb fields of surrounding plasma particles. These fields
modify the reaction rates and lead to
five nuclear burning regimes \cite{svh69} (two thermonuclear
regimes, with weak and strong plasma screening;
two pycnonuclear regimes, for zero-temperature and
thermally enhanced burning; and the intermediate regime). These  
regimes are described in Section \ref{fusion}; their validity
conditions are specified in Section \ref{phys}. In ordinary stars,
nuclear burning proceeds in the weak screening thermonuclear regime
\cite{FCZ,clayton83}.
The foundation of the
theory of thermonuclear burning with strong plasma screening
was laid by Salpeter \cite{Salp}. The strict theory of
pycnonuclear burning
was developed by Salpeter and Van Horn \cite{svh69}. References
to other works can be found in Section \ref{fusion}
and in \cite{leandro05}.
 
In Section~\ref{fusion} we analyze  calculations
of Coulomb barrier penetration in MCP
and propose a phenomenological expression for
a reaction rate valid in all five regimes for any non-resonant
fusion reaction. This expression accurately reproduces well
known results in thermonuclear regimes and 
gives a reasonable description of the Coulomb tunneling problem
in other regimes. It is important for incorporating the plasma 
physics effects into computer codes which simulate nucleosynthesis,
especially at high densities in compact stars, such
as white dwarfs and neutron stars (see above). 
In Section \ref{ignition} we illustrate the results of Sections
\ref{nuclear} and \ref{fusion}
by analyzing nuclear
burning in $^{12}$C\,$^{16}$O mixtures, with the emphasis
on the carbon ignition curve.

\section{Astrophysical $S$-factors for carbon-oxygen mixtures}
\label{nuclear}

\indent In order to study nuclear burning in dense stellar 
carbon-oxygen matter (Section \ref{ignition}) we need fusion cross sections 
(or associated astrophysical factors) for three reactions, 
$^{12}$C+$^{12}$C, 
$^{12}$C+$^{16}$O, and $^{16}$O+$^{16}$O. 
For calculating the cross sections we employ the
one-dimensional
barrier penetration (BP) formalism \cite{Fus04}
and adopt 
the S\~ao Paulo potential \cite{Rib97,Cha97,Cha98,Cha02}
to describe the real part of 
the nuclear interaction $V_N(r,E)$:
%
\begin{equation}
    V_N(r,E)=V_{SP}(r,E) = V_F(r) \; {\rm e}^{- 4 {\rm v}^2/c^2} \;.
\label{nuc_pot}
\end{equation}
Here, $V_F(r)$ is the density-dependent double-folding potential, $c$ is the
speed of light, $E$ is the particle collision energy (in the center-of-mass
reference frame), $\mbox{v}$ is the local relative velocity of two nuclei
1 and 2,
\begin{equation}
    \mbox{v}^2(r,E) = (2 /\mu) \,
    \left[ E-V_C(r)-V_{N}(r,E) \right] \;,
\label{speed}
\end{equation}
$V_C(r)$ is the Coulomb potential, and $\mu$ is the reduced mass.

In this paper, we adopt the two-parameter Fermi (2pF) distribution to describe 
the nuclear densities. The radii of these distributions 
are well approximated by 
the formula
$R_{0} = 1.31A^{1/3} - 0.84$ fm  \cite{Cha02}. 
The $^{12}$C and $^{16}$O diffuseness was taken to be $0.56$ fm 
and $0.58$ fm, respectively. 
These values 
were extracted from heavy-ion elastic 
scattering data at sub-barrier and intermediate energies, by applying an 
unfolding method involving the S\~ao Paulo potential 
(see Refs.~\cite{Alv01,Gas02,Ros02} for details).

\begin{figure*}[t]
\begin{center}
\includegraphics[width=16.0cm]{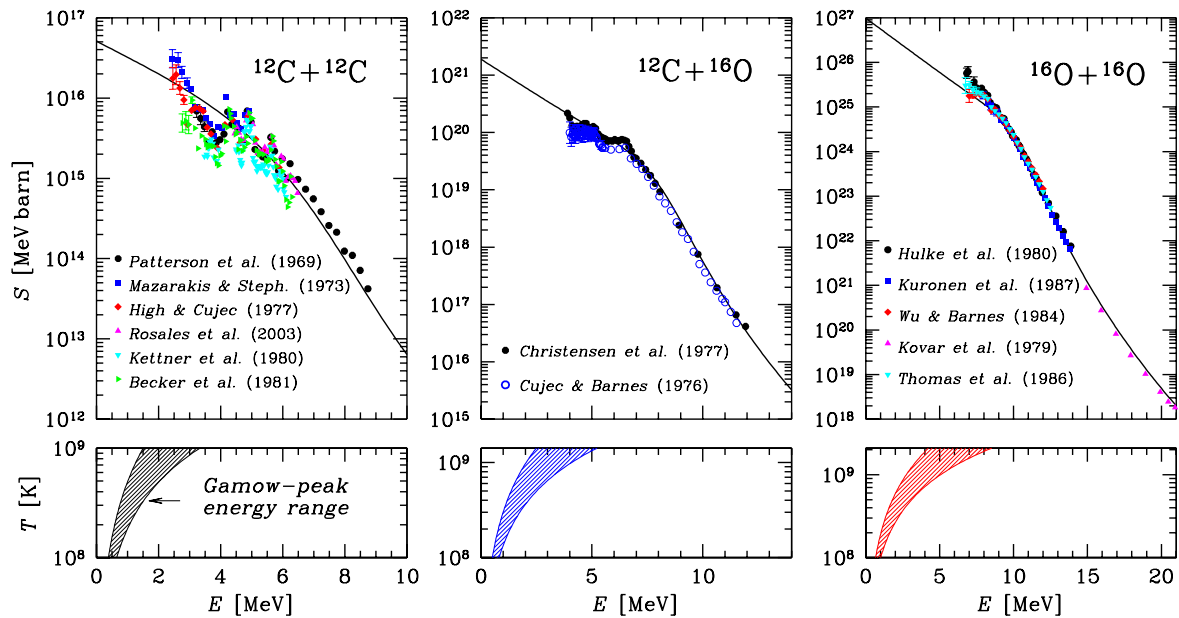}
\caption{(color online) 
{\it Top:} Astrophysical $S$-factors as a 
function of the
center-of-mass energy $E$ for the $^{12}$C+$^{12}$C,
$^{12}$C+$^{16}$O, and $^{16}$O+$^{16}$O reactions. 
The solid lines correspond to the BP model 
calculations while the various symbols are experimental data.
{\it Bottom:} Gamow-peak energy ranges for these reactions
in the thermonuclear regime versus the temperature $T$
of stellar matter. See text for details.}
\label{leandro}
\end{center}
\end{figure*}


Usually, fusion cross sections 
$\sigma(E)$ at low energies, typical for 
astrophysical conditions, are expressed in terms of the astrophysical 
$S$-factor
\begin{equation}
   S(E) = \sigma(E) \, E \; {\rm e}^{2\pi\eta} \ ,
\label{sf}
\end{equation}
where $\eta=(Z_1 Z_2 e^2/\hbar) \sqrt{\mu /(2E)}$ is the familiar Gamow 
parameter; $Z_1$ and $Z_2$ are charge numbers of the nuclei. This 
parameterization removes from the fusion cross section the strong 
non-nuclear energy 
dependence \cite{clayton83,barnes82} 
associated with Coulomb barrier penetration. 
If $S(E)$ is a slowly varying function of $E$, it can
be extrapolated to lower energies relevant to stellar burning. 

The $S$-factors for all three reactions versus $E$ are
shown in upper panels of Fig.\ \ref{leandro}. Solid lines
are theoretical calculations using the BP model, 
while symbols are experimental
data. The results for the C+C reaction have already been
discussed in Ref.\ \cite{leandro05} and are presented here
for completeness of consideration. The data for
this reaction are taken from  Patterson {\it et al.} \cite{Patterson},
Mazarakis and Stephens \cite{Mazarakis}, High and Cujec \cite{High},
Rosales {\it et al.} \cite{Rosales}, Kettner {\it et al.} \cite{Kettner},
and Becker {\it et al.} \cite{Becker}. The data for the
C+O reaction are taken from  Christensen {\it et al.} \cite{Chr77}
and Cujec and Barnes \cite{Cuj76}. Finally, the data for the O+O reaction
are from Hulke {\it et al.} \cite{Hul80}, Kuronen {\it et al.} \cite{Kur87},
Wu and Barnes \cite{Wu84}, Kovar {\it et al.} \cite{Kov79}, and
Thomas {\it et al.} \cite{Tho86}. Lower panels in Fig.\ \ref{leandro}
show the Gamow-peak energy ranges versus the temperature
of stellar matter in the thermonuclear burning regime;
they will be discussed in Section \ref{phys}.

The sub-Coulomb-barrier resonances exhibited in the 
C+C data at $E \lesssim 6$ MeV and in the
C+O data
at $E$ $\lesssim$ 7.7 MeV cannot be reproduced in the framework of
the BP model. 
However, the model provides an
average description of the light- and heavy-ion fusion at energies
below and above the barrier. Depending on the nuclear potential it
also gives a satisfactory parameter-free description of the
energy dependence of the $S$-factor,
which seems reasonably accurate 
for extrapolating the experimental data into the
stellar energy range. 

The data sets presented by the different groups are 
not in perfect agreement. For instance, 
the two C+O data sets agree in average, but    
disagree 
within a factor of $\sim2$ for lowest $E$. 
The overall agreement between the theory 
and the data is $\sim$50\%. However, in the 
low-energy region, the slope of the 
measured cross section reported in Ref. \cite{Chr77} does not follow 
the calculated $S$-factors. It is difficult to predict where the data will lie 
in the energy range $E\lesssim 4$ MeV. 
Regarding the O+O reaction, 
the discrepancies between the different experimental results at sub-barrier 
energies are around a factor of 3. 
The overall agreement between 
the data and the theory
is $\sim$30\%.
At the lowest measured energy the difference between the data 
and the theory are at most a factor of 3. 
Further experiments at lower
energies would help in verifying the validity of the predicted
fusion cross sections. Nevertheless, it is important to highlight
that the BP model does not contain any free parameter. In this
sense, the $S$-factor calculations do not represent a
fit to the experimental data and can be considered as a useful tool
to predict average non-resonant low-energy cross sections for a
wide range of fusion reactions. For many astrophysical reactions,
such a description gives a reasonable estimate because the
formalism of stellar reaction rates relies on the $S$-factor
averaged over an entire Gamow-peak range.

The values of $S(E)$ calculated up to 
$E \leq 20$ MeV can be fitted by an analytic expression
\begin{eqnarray}
   S(E)&=&
   \exp \left( A_1+A_2\,\Delta E  {{}\over{}}  \right. 
\nonumber \\   
   & +&  
   \frac{A_3+A_4\,\Delta E+A_5\,\Delta E^2}{1+{\rm e}^{-\Delta E}}
  \left. {{}\over {}} \right)~~{\rm MeV~barn},
\label{fit}
\end{eqnarray}
where $\Delta E=E-E_0$; the center-of-mass energy $E$ and the fit parameter 
$E_0$ are expressed in megaelectron-volts. Table \ref{tab:sfac} gives the 
fit parameters $A_1$, \ldots, $A_5$ and $E_0$ for the 
C+C, C+O and O+O reactions. The 
maximum formal fit errors are 7.2\% at $E$ = 19.8 Mev for C+C; 6.3\% at 
$E$ = 7.5 MeV for C+O; and 3.9\% at $E$ = 8.2 MeV for the O+O reaction. The 
$S$-factor for the C+C reaction was fitted in a previous paper \cite{leandro05}
by a slightly different expression with approximately the same accuracy. We 
have fitted the same data by the new expression (\ref{fit}) for completeness 
of consideration. The two fits are nearly equivalent.

\begin{table}[t]
\caption[]{The coefficients $A_1$,\ldots,$A_5$ and 
$E_0$ in the fits expression 
(\ref{fit}) for the $S$-factors of the C+C, C+O and O+O reactions.}
\label{tab:sfac}
\begin{center}
\begin{tabular}{c c c c c c c }
\hline
\hline
~Reaction~& ~~~$E_0$~~~  & ~~~$A_1$~~~  &~~~ $A_2$~~~ &  ~~~$A_3$~~~ &
~~~~$A_4$~~~~~~  & ~$A_5$ \\
\hline 
$^{12}$C+$^{12}$C & 6.946 & 34.75 & $-0.552$ & $-2.131$ &
$-0.625$ & 0.0315  \\
$^{12}$C+$^{16}$O & 8.290 & 44.32 & $-0.561$ & $-1.480$ &
$-0.910$ & 0.0387  \\
$^{16}$O+$^{16}$O & 10.52 & 56.16 & $-0.571$ & $-1.160$ &
$-1.044$ & 0.0366  \\
\hline
\hline
\end{tabular}
\end{center}
\end{table}

\section{Nuclear fusion rate}
\label{fusion}

\subsection{Physical conditions and reaction regimes}
\label{phys}

Let us consider a stellar matter which consists of
ions and electrons. We assume that the ions are fully ionized
and the electrons form a uniform electron background.
We study a multi-component mixture of ion species
$j=1,2,\ldots$, with atomic numbers $A_j$ and charge numbers
$Z_j$. Let $n_j$ be the number density of ions $j$.
The total number density of ions is $n=\sum_j n_j$; the
electron number density is $n_e = \sum_j Z_j n_j$.
For an OCP of ions the index $j$ will
be omitted. The number density $n_j$ can be expressed through the
mass density $\rho$ of the matter as $n_j=X_j \rho/(A_j m_{\rm u})$,
where $X_j$ is the mass fraction of ions $j$,
and $m_{\rm u}=1.66055 \times 10^{-24}$~g is the atomic mass unit.
In a not too dense matter the total mass
fraction contained in the nuclei is $X_N = \sum_j X_j \approx 1$. At densities
higher than the neutron drip density
($\sim 4 \times 10^{11}$~g~cm$^{-3}$), the matter contains
also free neutrons; the total mass fraction contained
in the nuclei is then $X_N<1$.
It is also useful to introduce
the fractional number $x_j=n_j/n$ of nuclei $j$ among other nuclei,
with $\sum_j x_j=1$.
Generally,
\begin{eqnarray}
&&     n_e = n \langle Z \rangle, \quad
     \rho={m_{\rm u} n \langle A \rangle \over X_N}, 
      \quad     
     x_j = { X_j/A_j \over \sum_i X_i/A_i},
\nonumber \\
&&     \langle Z \rangle = \sum_j x_j Z_j, \quad
     \langle A \rangle = \sum_j x_j A_j, 
\label{ne-rho}
\end{eqnarray}
where $\langle Z \rangle$ and $\langle A \rangle$ are the mean
charge and mass number of ions, respectively.

\begin{figure}[t]
\begin{center}
\epsfysize=80mm
\epsffile[20 165 465 585]{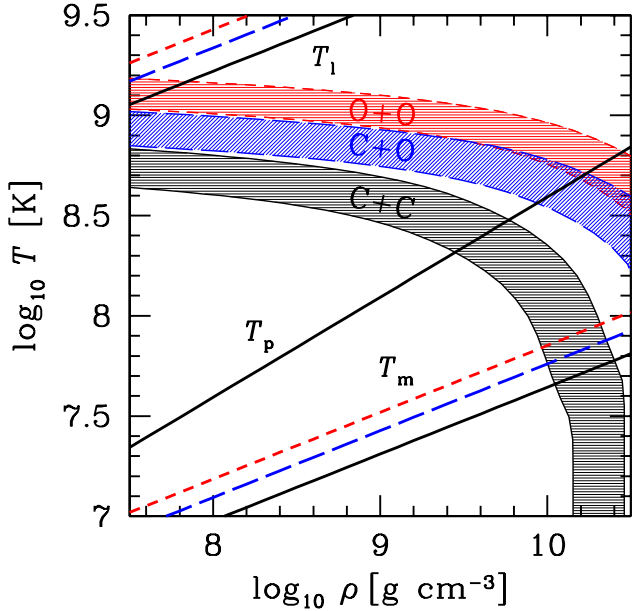}
\caption{(color online) Temperature-density diagram for a 
$^{12}$C\,$^{16}$O matter.
Straight lines show	
the temperature $T_l$ of the appearance of ion
liquid, the melting temperature  $T_{m}$ of ion
crystal, and the ion plasma temperature $T_p$.
Solid lines refer to pure carbon matter,
long dashes to CO matter with equal
particle fractions of C and O, and short dashes to
pure oxygen matter. Three shaded strips
show the regions important for C+C burning (in pure
carbon matter), C+O burning (in CO mixture) and
O+O burning (in pure oxygen matter).
Each strip is restricted by upper and lower
lines along which the
burning time equals 1 year and $10^6$ years, respectively (see the text
for details).}
\label{fig:diag}
\end{center}
\end{figure}

Let us also introduce the Coulomb coupling parameter $\Gamma_j$
for ions $j$,
\begin{eqnarray}
 &&  \Gamma_j={ Z_j^2 e^2 \over a_j k_{\rm B} T}
   ={Z_j^{5/3}e^2 \over a_e k_{\rm B} T} ,
\label{Gammaj} \\   
 &&  a_e=\left( 3 \over 4 \pi n_e \right)^{1/3},
   \quad
   a_j=Z_j^{1/3}a_e,
\nonumber    
\end{eqnarray}
where $T$ is the temperature,
$k_{\rm B}$ is the Boltzmann constant, $a_e$ is the electron-sphere
radius, and
$a_j$ is the ion-sphere radius (a radius of a
sphere around a given ion, where the electron charge compensates
the ion charge). Therefore, $\Gamma_j$ is the ratio
of a typical electrostatic energy of the ion to
the thermal energy.
If $\Gamma_j \ll 1$ then the ions
constitute an almost ideal Boltzmann gas, while for $\Gamma_j \gtrsim 1$
they are strongly coupled by Coulomb forces (constitute either
Coulomb liquid or solid). The transformation from the gas to the liquid
at $\Gamma_j \sim 1$ is smooth, without any phase transition.
The solidification is realized as a weak second-order phase
transition. According to highly accurate Monte Carlo
calculations, a classical OCP of ions
solidifies at $\Gamma \approx 175$
(see, e.g., Ref.~\cite{dewittetal03}).

It is useful to introduce the mean ion coupling parameter
$\langle \Gamma \rangle = \sum_j x_j \Gamma_j$. The
plasma can be treated as strongly coupled 
if $\langle \Gamma \rangle \gtrsim 1$.
This happens at $T \lesssim T_l$, where
\begin{equation}
   k_{\rm B}\,T_l=  \sum_j (Z_j^2 e^2/ a_j)\,x_j= 
   k_{\rm B}T \langle \Gamma \rangle.
\label{Tl}
\end{equation}

For low temperatures $T \ll T_p \ll T_l$ ion motion cannot
be any longer considered as classical but should be quantized.
Here, $T_p$ is the Debye (plasma) temperature associated with
the ion plasma frequency $\omega_p$ (a typical
frequency of ion vibrations in Coulomb crystals -- see, e.g., 
Ref.\ \cite{svh69}),
\begin{equation}
    T_p = {\hbar \omega_p \over k_{\rm B}},
    \quad
    \omega_p^2=  \sum_j { 4 \pi Z_j^2 e^2 n_j
      \over A_jm_{\rm u}} .
\label{Tp}
\end{equation}

The most difficult problem of a strongly coupled MCP
at low temperatures consists in understanding
its actual state. Extensive Monte Carlo simulations \cite{dsy93}  
of the freezing of a classical OCP 
indicate that it can freeze into imperfect 
body-centered cubic (bcc) or
faced-centered cubic (fcc) microcrystal (or microcrystals).
Unfortunately, publications on reliable simulations
of freezing of an MCP are almost absent. 
Evidently, the cold MCP is much more rich in physics than
the OCP. It can be an MCP regular lattice or
microcrystals (with defects); 
or an amorphous, uniformly mixed structure;
or a lattice of one phase with random admixture
of other ions. One cannot exclude an ensemble
of phase separated domains. For the sake of completeness,
we will consider different possibilities.
An MCP obeys the linear mixing rule with the high accuracy.
Accordingly, the difference in energies of the indicated states
is very small and is a subject of vigorous debates
(e.g., \cite{ds03} and references therein).
It is possible that a low-temperature MCP can be in different states
depending on the history of its formation in a star with
decreasing temperature.

To make our consideration less abstract,
we will apply it to a 
$^{12}$C\,$^{16}$O mixture. The appropriate temperature-density
diagram is shown in Fig.~\ref{fig:diag}.
We present the temperatures $T_{l}$, $T_{m}$ and
$T_{p}$ for a pure carbon matter ($x_{\rm C}=1$, solid lines),
for a mixture of equal amounts of C and O nuclei
($x_{\rm C}=\frac12$, long dashes) and for a pure oxygen matter
($x_{\rm C}=0$, short dashes). The melting temperature of the CO mixture
is taken to be $T_{m}=T_{l}/175$. The plasma temperature
$T_{p}$ is the same for all three cases. Notice
that the electrons are strongly degenerate at all
$\rho$ and $T$ displayed in Fig.~\ref{fig:diag}.
At $\rho>4 \times 10^{10}$ g~cm$^{-3}$
carbon nuclei cannot survive in dense matter because
of beta captures; at $\rho> 2 \times 10^{10}$ g~cm$^{-3}$
oxygen nuclei will also be destroyed by beta captures.
Therefore, it is unreasonable to extend the CO diagram to higher densities.

We will be interested in nuclear fusion reactions
\begin{equation}
    (A_i,Z_i)+(A_j,Z_j) \to (A_c,Z_c),
\label{reaction}
\end{equation}
where $A_c=A_i+A_j$ and $Z_c=Z_i+Z_j$ refer to a compound nucleus
$c$. For our example in Fig.~\ref{fig:diag}, we have three
reactions, C+C, C+O,
and O+O (Section \ref{nuclear}).
The experimental cross sections for these reactions show
no very pronounced resonance structures  
and can be described in the framework of non-resonant
reaction formalism \cite{barnes82,leandro05} as discussed above.
Notice, that the data for the C+C and C+O reactions exhibit
some resonant structures demonstrated in Fig.\ \ref{leandro}.
In the lower panel of this figure we show the Gamow-peak
energy range as a function of temperature for these reactions
in the thermonuclear burning regime (Section \ref{thermo}).
We see that at high enough $T$ the Gamow-peak range covers
the energy range, where the oscillations are experimentally measured.
However, at these values of $T$ the Gamow-peak range is sufficiently
wide and the energy integration in the reaction rate should
smear out the oscillatory behavior. For lower $T$ the Gamow
peak is narrower, but it shifts to low energies inaccessible
to laboratory experiments. In the absence of experimental
and theoretical data on the presence of oscillations at these
low energies we will adopt the standard assumption that the
reactions in question can be treated as non-resonant in 
applications to stellar burning.

The shaded strips Fig.\ \ref{fig:diag}
show the $T-\rho$ domains most important for these
reactions. We will describe them in more
detail in Section \ref{ignition}. 

To study a reaction (\ref{reaction}) we introduce
the so called {\it ion-sphere quantities}
\begin{eqnarray}
  && a_{ij}={a_i+a_j \over 2}, \quad
   \Gamma_{ij}={Z_i Z_j e^2 \over a_{ij}k_{\rm B}T}, \quad
   T_{ij}^{(l)}={Z_iZ_j e^2 \over a_{ij} k_{\rm B}}, 
\nonumber \\   
 &&  T_{ij}^{(p)}= {\hbar \over k_{\rm B}} \,
   \left( 4 \pi Z_i Z_j e^2 n_{ij} \over 2 \mu_{ij} \right)^{1/2},
\label{notations}
\end{eqnarray}
where $\mu_{ij}=m_{\rm u}A_i A_j/A_c$ is the reduced mass
of the reacting nuclei, and $n_{ij}=3/(4 \pi a_{ij}^3)$.
Basing on the ion-sphere model of a strongly coupled Coulomb plasma,
one expects
(e.g., Ref.\ \cite{ichimaru82}) that $a_{ij}$ characterizes an equilibrium distance
between neighboring nuclei $i$ and $j$, $\Gamma_{ij}$ 
describes
their Coulomb coupling, $T_{ij}^{(l)}$ is
the temperature for the onset of strong coupling,
and $T_{ij}^{(p)}$ is a
local Debye temperature (for
oscillations of ions $i$ and $j$). 
In an OCP we have $a_{ij}=a$,
$\Gamma_{ij}=\Gamma$, $T_{ij}=T_p$. We will also need
\begin{equation}
     r_{{\rm B}ij}={\hbar^2 /( 2 \mu_{ij}   Z_i Z_j e^2)},
\label{Bohr}
\end{equation}
which reduces to the ion Bohr radius in the case of equal ions $j=i$.
In addition, we will need the parameter
\begin{eqnarray}
    \lambda_{ij}&=&r_{{\rm B}ij} \, \left( n_{ij} \over 2 \right)^{1/3}=
    {2 r_{{\rm B}ij} \over (Z_i^{1/3} + Z_j^{1/3}) }
    \left( \rho X_N \langle Z \rangle
    \over 2 \langle  A \rangle m_{\rm u} \right)^{1/3}
\nonumber \\
    &=& {A_i+A_j \over A_i A_j Z_i Z_j (Z_i^{1/3} + Z_j^{1/3})  }
\nonumber \\
   && \times \left( \rho X_N \langle Z \rangle \over
    \langle A \rangle \;
    1.3574 \times 10^{11}~{\rm g~cm}^{-3} \right)^{1/3},
\label{lambda}
\end{eqnarray}
which corresponds to the parameter $\lambda$ introduced by
Salpeter and Van Horn \cite{svh69} in the OCP case.

In the following sections we will discuss nuclear burning in
MCP for the five burning regimes introduced in Ref.\ \cite{svh69}
and analyzed in detail for OCP in our previous work
\cite{leandro05}. We will demonstrate that the
formalism developed for OCP can be adapted to
more complex MCP scenarios.

\subsection{Classical thermonuclear reaction rate}
\label{thermo}

\indent In the classical thermonuclear (weak screening)
regime ($T \gg T_{ij}^{(l)}$) the reacting ions constitute
an almost ideal gas \cite{FCZ}.
The rate for non-resonant fusion processes (such as
considered here) is well known,
\begin{equation}
    R_{ij}^{\rm th}= {4\, n_i \, n_j \over 1 + \delta_{ij}} \,
    \sqrt{ 2 E_{ij}^{\rm pk} \over 3 \mu_{ij}} \,
    { S(E_{ij}^{\rm pk}) \over k_{\rm B} T } \, \exp(-\tau_{ij}),
\label{therm}
\end{equation}
where $S(E)$ is the astrophysical factor;
$\delta_{ij}$ is the Kronecker delta, which excludes
double counting of the same collisions in reactions with
identical nuclei ($i=j$); $E_{ij}^{\rm pk}=T k_{\rm B} \tau_{ij} /3$ is the
Gamow peak energy (the relative energy of colliding nuclei
which gives the major contribution into the reaction rate) and
\begin{equation}
  \tau_{ij}= \left( 27 \pi^2 \mu_{ij}\, Z_i^2\, Z_j^2 e^4 \over
  2 k_{\rm B} T \hbar^2 \right)^{1/3}
\label{tau}
\end{equation}
is the parameter which characterizes the
Coulomb barrier penetrability. 
This parameter can be written as
\begin{equation}
    \tau_{ij}=3\,(\pi/2)^{2/3}(E_{ij}^{\rm a}/k_{\rm B} T)^{1/3}, \quad
    E^{\rm a}_{ij} \equiv 2 \mu_{ij} Z_i^2 Z_j^2 e^4/\hbar^2.
\label{tau1}
\end{equation}

Then
\begin{equation}
    R_{ij}^{\rm th}= { n_i n_j \over 1 + \delta_{ij}}
      \, S(E_{ij}^{\rm pk}) \,
     { r_{{\rm B}ij} \over \hbar }\, P_{\rm th} \, F_{\rm th},
\label{Rth}
\end{equation}
where $r_{{\rm B}ij}$ is a convenient dimensional
factor defined by Eq.~(\ref{Bohr}), $F_{\rm th}$ is the exponential
function, and $P_{\rm th}$ is the pre-exponent,
\begin{equation}
     F_{\rm th}=\exp(-\tau_{ij}), \quad
     P_{\rm th}= { 8 \pi^{1/3} \over \sqrt{3}\, 2^{1/3}} \,
       \left(E_{ij}^{\rm a} \over k_{\rm B} T \right)^{2/3}.
\label{Exth}
\end{equation}

Typically, the main contribution into the reaction rate
comes from suprathermal particles ($E_{ij}^{\rm pk} \gg k_{\rm B} T$),
and the Coulomb barrier is very thick ($\tau_{ij} \gg 1$).
The reaction rate decreases exponentially
with decreasing $T$.
Typical Gamow-peak energy ranges for the C+C, C+O, and O+O
reactions in the thermonuclear regimes are shown in
Fig.\ \ref{leandro}. These energies are defined as
$E_{ij}^{\rm pk}-\Delta E \lesssim E \lesssim E_{ij}^{\rm pk}+\Delta E$,
with $\Delta E \sim 2 \sqrt{ E_{ij}^{\rm pk} k_{\rm B}T}$.

\subsection{Thermonuclear regime with strong screening}
\label{thermoscreen}

\indent The thermonuclear regime with
strong plasma screening operates in the temperature
range $T_{ij}^{(p)} \lesssim T \lesssim T_{ij}^{(l)}$,
where the plasma ions constitute a strongly coupled Coulomb system
(liquid or solid).
The majority of ions are confined in deep Coulomb potential wells
but the main contribution into the reaction rate comes from
a very small amount of highly energetic suprathermal ions
which are nearly free (see, e.g., Refs.\
\cite{Salp,svh69}). However, neighboring plasma ions
strongly screen the Coulomb interaction between the reacting
ions. The screening simplifies close approaches of the reacting
ions, required for a Coulomb tunneling,
and enhances thus the reaction rate (with respect to the classical
thermonuclear reaction rate).

In analogy with the OCP case
(e.g., Ref.\ \cite{leandro05}), the enhancement can be included
into the exponential function $F_{\rm th}$,
\begin{equation}
     F_{\rm th}=F_{\rm sc}\,\exp(-\tau_{ij}),
     \quad F_{\rm sc}=\exp(h_{ij}),
\label{Exscr}
\end{equation}
where $F_{\rm sc}$ is the enhancement factor and $h_{ij}$ is a function of
plasma parameters.

We will analyze $h_{ij}$ in the same manner as was
done in Ref.\ \cite{leandro05} for the OCP.
For this purpose we notice that the reacting nuclei move in the potential
$W(r)=Z_iZ_je^2/r-H_{ij}(r)$, where $H_{ij}(r)$ is the plasma
potential created by neighboring plasma ions.
In the thermonuclear
regime, $H_{ij}(r)$ is almost constant along a Coulomb tunneling
path. Accordingly, $h_{ij}$ can be split into two terms,
$h_{ij}=h_{ij}^{(0)}+h_{ij}^{(1)}$. The main term $h_{ij}^{(0)}$
is obtained assuming that 
$H_{ij}(r)\approx H_{ij}(0)$ is constant along a tunneling path;
a small correction $h_{ij}^{(1)}$ is produced by a weak variation of
the plasma potential along this path. We will discuss $h_{ij}^{(0)}$
explicitly in this section and introduce $h_{ij}^{(1)}$ phenomenologically
in Section \ref{together}, when we propose an analytic approximation for the
reaction rate in all regimes.

It is well known (e.g., Refs.\ \cite{Salp,dgc73})
that $h_{ij}^{(0)}={H}_{ij}(0)/k_{\rm B}T$,
where $H_{ij}(0)$ is the properly averaged (mean-field) plasma potential
at $r=0$. Thus defined, $H_{ij}(0)$ becomes a thermodynamic
quantity which can be expressed through the difference of
the classical Coulomb free energies for a given system and a system with the
two reacting nuclei merging into a compound nucleus (e.g., Ref.\ \cite{dgc73}).
However, a strongly coupled classical multi-component
Coulomb liquid obeys very
accurately the linear mixing rule (see Ref.\ \cite{ds03}
for recent results).
Using this rule, one obtains
\begin{equation}
  h_{ij}^{(0)}=f_0(\Gamma_i)+f_0(\Gamma_j)-f_0(\Gamma_c),
\label{jancovici}
\end{equation}
where $f_0(\Gamma)$ is the Coulomb free energy per ion in an OCP
(in units of $k_{\rm B}T$).
This formula was derived by Jancovici
\cite{jancovici77} for an OCP and generalized by Mochkovitch
for an MCP (see Ref.~\cite{mn86}).
The function $f_0(\Gamma)$
is very accurately determined by Monte Carlo simulations.
For instance, according to
DeWitt and Slattery \cite{ds99}, the function $f_0(\Gamma)$ for a classical
one-component Coulomb liquid at $1 \leq \Gamma \leq 170$ can be
approximated as
\begin{eqnarray}
 f_0(\Gamma)& =& -0.899172 \, \Gamma + (1/s) \, 0.602249 \,
              \Gamma^s
\nonumber \\	       
	   &&   - 0.274823 \, \ln \Gamma
  - 1.401915,
\label{strong}
\end{eqnarray}
where $s=0.3230064$. It gives
a highly accurate expression for $h_{ij}^{(0)}$, but it is
inconvenient for an analytic interpolation
of the reaction rate  (Section
\ref{together}). Instead, we will use a 
simpler linear expression $f_0(\Gamma)=-0.9\,\Gamma$ provided
by the ion-sphere model \cite{Salp},
\begin{eqnarray}
 &&    h_{ij}^{(0)}=C_{ij}^{\rm sc}\,\Gamma_{ij},
\label{salpeter}\\     
 &&    C_{ij}^{\rm sc}=0.9 \,\left[ Z_c^{5/3}-Z_i^{5/3}-Z_j^{5/3} \right]
     \,{Z_i^{1/3}+Z_j^{1/3} \over 2Z_iZ_j}.
\nonumber
\end{eqnarray}
This expression seems crude but, actually, it is accurate.
For a nuclear reaction in an OCP
($Z_i=Z_j=Z_c/2$), this equation
gives $C^{\rm sc}=1.0573$, very close to 
the value 1.0754 inferred \cite{leandro05} from Eq.\ (\ref{strong}).
In the range of $1 \leq \Gamma_{ij} \leq 170$ the ion-sphere model
(\ref{salpeter}) gives the enhancement factor $\exp(h_{ij}^{(0)})$
which is systematically lower than the more accurate enhancement factor,
given by Eqs.~(\ref{jancovici}) and (\ref{strong}).
The maximum difference of these enhancement factors
for charge ratios $1/5 \leq Z_i/Z_j \leq 5$
reaches $\approx 15$ at the highest value of $\Gamma_{ij}=170$.
This difference can be regarded as insignificant
because at such $\Gamma_{ij}$
the enhancement factor itself is as huge as
$\exp(h_{ij}^{(0)}) \sim 10^{74}$.
For lower $\Gamma_{ij}$ the expression (\ref{salpeter})
is more accurate. For instance, for $\Gamma_{ij}=50$ 
and $Z_i/Z_j=5$ it underestimates 
the enhancement factor only by a factor of 3. In the range of
$1 \leq \Gamma_{ij} \leq 10$ for  $1/5 \leq Z_i/Z_j \leq 5$
the underestimation does not exceed a factor of 1.5.
Notice that in the MCP the factor $C^{\rm sc}_{ij}$ depends
on $Z_i$ and $Z_j$.

Although the above approach is more accurate,
one usually calculates $h_{ij}^{(0)}$ 
by extrapolating the mean-field plasma
potential $H_{ij}(r)$, obtained from classical
Monte Carlo sampling, to $r \to 0$. In particular, Ogata {\it et al.}
\cite{oii91,oiv93} used this method to study the enhancement
of thermonuclear burning in the liquid phases of  OCP
and binary ion mixtures (BIMs).
The leading term in Eqs.~(19) and (20)
of Ref.~\cite{oiv93}, equivalent to the leading terms
(\ref{jancovici}) and (\ref{salpeter}), is
$h_{ij}^{(0)}=\Gamma_{ij}\, (1.148 - 0.00944\, \ln \Gamma_{ij}
-0.000168\,(\ln \Gamma_{ij})^2)$.
This expression was employed also by Kitamura \cite{kitamura00}
for constructing the analytic approximation for the reaction rates
in OCP and BIMs
in all reaction regimes (although by that time a more
accurate expression was obtained by Ogata
\cite{ogata97} for the OCP using path integral Monte Carlo simulations).
Comparing the expression of Ref.~\cite{oiv93} with
(\ref{jancovici}) and (\ref{salpeter})
we see that the expression of Ref.~\cite{oiv93}
systematically overestimates
the plasma enhancement.
In the OCP the overestimation reaches \cite{leandro05}
a factor of $\sim 40$ for $\Gamma=170$, which is not very significant.
However, the coefficients in this expression are independent of
$Z_i$ and $Z_j$, in disagreement with Eq.\
(\ref{salpeter}). As a result, the overestimation increases
with the growth of the charge ratio $Z_i/Z_j$, reaching $\sim 150$ 
and $\sim 2 \times 10^4$ at
$Z_i/Z_j=2$ and $Z_i/Z_j=5$, respectively,
for $\Gamma_{ij}=170$. Therefore,
when the difference of charges $Z_i$ and $Z_j$ increases,
the results of Refs.~\cite{oiv93} and \cite{kitamura00} become
less accurate. The nature of inaccuracy comes from the problems
of extrapolation of $H_{ij}(r)$ to $r \to 0$. It was analyzed
by Rosenfeld for the OCP
\cite{rosenfeld96} and the MCP \cite{rosenfeld96b}.

Equations (\ref{strong}) and (\ref{salpeter}) become invalid
in the regime of weak screening ($\Gamma_{ij} \ll 1$;
Section \ref{thermo}), where
the well known Debye-H\"uckel
theory should be used. In particular, the screening function in the
weakly coupled MCP becomes \cite{Salp}
\begin{equation}
    h_{ij}^{(0)}= { Z_i Z_j e^2 \over k_{\rm B} T r_{\rm D}}
    =\left( 3 \Gamma_e^3 \langle Z^2 \rangle Z_i^2 Z_j^2 \over
    \langle Z \rangle \right)^{1/2},    
\label{DH}
\end{equation}
where $r_{\rm D}$ is the ion Debye screening length and
$\langle Z^2 \rangle \equiv  \sum_j Z_j^2 x_j$.
Introducing $\Gamma_{ij}$, we obtain
\begin{equation}
   h_{ij}^{(0)}= \zeta_{ij} \, \Gamma_{ij}^{3/2}, \qquad
   \zeta_{ij}=\left(  3 \langle Z^2 \rangle (Z_i^{1/3}+Z_j^{1/3})^3
   \over 8 \langle Z \rangle Z_i Z_j \right)^{1/2}.
\label{DH1}
\end{equation}
For reactions in an OCP, we have $\zeta=\sqrt{3}$.
In an MCP, $\zeta_{ij}$ depends on ion charge numbers.

A simple phenomenological interpolation 
which reproduces the strong and weak screening
limits (Eqs.\ (\ref{salpeter}) and (\ref{DH1})) 
and combines them in the
$\Gamma_{ij}$ range from $\sim 0.1$ to $\sim 1$ is
\begin{equation}
     h_{ij}^{(0)}=C^{\rm sc}_{ij}\, \Gamma_{ij}^{3/2}/
    [(C_{ij}^{\rm sc}/\zeta_{ij})^4+\Gamma_{ij}^2]^{1/4}.
\label{scrfit}
\end{equation}
Because accurate
calculations of the MCP free energy of ions  
in this range
are absent, we cannot test the accuracy of our interpolation.
However, the plasma screening enhancement
of reaction rates at these values of $\Gamma_{ij}$ is weak and  
the interpolation uncertainty does not affect strongly 
the reaction rates.

\subsection{Zero-temperature pycnonuclear regime}
\label{pycnozero}

This regime operates at low temperatures,
$ T \lesssim 0.5\, T_{ij}^{(p)}/ \ln(T_{ij}^{(l)}/T_{ij}^{(p)})$,
at which thermal effects are negligible and all the ions occupy
ground states in their potential wells. The Coulomb barrier
is penetrated
owing to zero-point vibrations of ions around their equilibrium
positions. Because the vibration amplitudes are generally
small, neighboring pairs of ions
(closest neighbors) make the major contribution 
into the reaction rate.

Generalizing Eq.~(35) of Salpeter and Van Horn \cite{svh69}
to the MCP case we can
present the pycnonuclear reaction rate as
(see, e.g., Eq.~(7) in Ref.~\cite{oii91})
\begin{equation}
   R_{ij}^{\rm pyc} = {n_i \over 1 + \delta_{ij}} \,
   \langle \nu_{ij}\, p_{ij} \rangle_{\rm av},
\label{pyc1}
\end{equation}
where $\nu_{ij}$ is the number of nearest nuclei $j$
around a nucleus $i$, $p_{ij}$ is the reaction rate for
a fixed pair $ij$, and the brackets $\langle \ldots \rangle_{\rm av}$
denote statistical averaging over an ensemble of such pairs.
For instance, it is currently thought that the OCP
of ions at zero temperature forms a 
bcc
crystal. In this case, any ion is surrounded by 8 closest
neighbors, with the
equilibrium distance between them
$d=(3 \pi^2)^{1/6} a$, where $a$ is the
ion-sphere radius.

According to Eqs.\ (35), (37), and (39) of Ref.\ \cite{svh69},
the reaction rate for a pair of neighboring
ions in an OCP
can be written as
\begin{equation}
    p= D_{\rm pyc} \,
    {\lambda^{3-C_{\rm pl}} \,S(E^{\rm pk}) \over \hbar r_{\rm B}^2 }
    \, \exp \left( - {C_{\rm exp} \over \sqrt{\lambda} }   \right),
\label{svh1}
\end{equation}
where $\lambda$ and $r_{\rm B}$ are given by Eqs.\ (\ref{lambda})
and (\ref{Bohr}), respectively (for the OCP);
while $D_{\rm pk}$, $C_{\rm pl}$ and $C_{\rm exp}$ are constants
which depend on a Coulomb barrier penetration model and
on the lattice type. Finally, the characteristic reaction energy
is $E^{\rm pk} \sim \hbar \omega_p =k_{\rm B}T_p$,
where $\omega_p$ is the
ion plasma frequency defined by Eq.~(\ref{Tp})
(a typical frequency of zero-point ion vibrations).

Salpeter and Van Horn \cite{svh69} used the
WKB approximation and considered
two models of Coulomb barrier penetration
in the bcc lattice, the static and
relaxed lattice ones, to account for the lattice response to
a motion of tunneling nuclei.
Later Schramm and Koonin \cite{Schramm} extended this consideration
taking into account the
dynamical effect of motion of surrounding ions in response to the
motion of the tunneling nuclei in the relaxed lattice.
In addition, they
considered the 
fcc
Coulomb lattice.
The results of Refs.\ \cite{svh69,Schramm} are analyzed
in Ref.~\cite{leandro05}. Note that the pycnonuclear burning
rates for fcc or bcc OCP crystals are very similar. We expect that
for an amorphous OCP they are of the same
order of magnitude.

Pycnonuclear reactions in the MCP
require complicated calculations (which, hopefully, will
be done in the future). We will restrict ourselves to a simpler
consideration based on similarity criteria
and some general assumptions.
Even a state of the MCP
at low temperatures
is not clear. It can be 
a regular lattice (with defects); a uniformly mixed state;
an ensemble of phase-separated domains, etc.\ (Section \ref{phys}). 
It can also be a combination
of these states.

\subsubsection{Uniformly mixed MCP}
\label{umix}

Let us start with 
a uniformly mixed MCP.
An obvious generalization of Eq.\ (\ref{svh1}) to the MCP
would be to replace $r_{\rm B} \to r_{{\rm B}ij}$ and
$\lambda \to \lambda_{ij}$ in accordance with Eqs.\ (\ref{Bohr})
and (\ref{lambda}). This replacement should correctly reflect the
rescaling of inter-ion distances 
and oscillator frequencies in an MCP
within the ion-sphere model (see, e.g., Ref.~\cite{iovh92}).

However, the rescaling may be not exact.
We will take this into account by
adopting a simplified assumption that any pair of close neighbors
behaves as {\it an elementary oscillator} with an
equilibrium separation $d_{ij}$ and an oscillator frequency
$\omega_{ij}$.
For an OCP we have $\lambda \sim \hbar^2/
(A m_{\rm u} d^2 \omega)^2$ and $\omega \sim \omega_p$.
In the MCP we expect to have
$\lambda \to
\widetilde{\lambda}_{ij}
\sim \hbar^2/(2 \mu_{ij} d_{ij}^2 \omega_{ij})^2$,
where $d_{ij}$ and $\omega_{ij}$ are the actual equilibrium distance
and the effective oscillator frequency, respectively. With this replacement
from Eq.\ (\ref{svh1}) we obtain
\begin{equation}
    p_{ij}= D_{\rm pyk} \,
    {\widetilde{\lambda}_{ij}^{3-C_{\rm pl}} \,S(E_{ij}^{\rm pk})
    \over \hbar r_{{\rm B}ij}^2 }
    \, \exp \left( - {C_{\rm exp} \over
    (\widetilde{\lambda}_{ij})^{1/2} }  \right).
\label{svh2}
\end{equation}

Now we assume that the actual values
$d_{ij}$ and $\omega_{ij}$ can deviate from the ion-sphere values
$d_{ij}^{(0)}$ and $\omega_{ij}^{(0)}$, and
introduce the quantities
\begin{equation}
     \alpha_{{\rm d}ij}=d_{ij}/d_{ij}^{(0)}, \qquad
     \alpha_{\omega ij}=\omega_{ij}/\omega_{ij}^{(0)},
\label{alphas}
\end{equation}
which measure the deviations. They will be treated
as parameters to be varied within reasonable limits. 
Then
\begin{equation}
     \widetilde{\lambda}_{ij}=\lambda_{ij} \, \alpha_{\lambda ij}, \qquad
     \alpha_{\lambda ij}\equiv 1/(\alpha_{{\rm d}ij}^4 \alpha_{\omega ij}^2),
\label{lambda1}
\end{equation}
where $\lambda_{ij}$ is given by Eq.\ (\ref{lambda}).

Another way to improve the ion-sphere rescaling
for a pycnonuclear reaction in a BIM was proposed
by Ichimaru {\it et al}.~\cite{iovh92} who used the
formal relation $\lambda_{ij} \sim r_{{\rm B}ij}/d_{ij}$
and suggested that
$\widetilde{\lambda}_{ij}=\lambda_{ij}/\alpha_{{\rm d}ij}$,
which corresponds to $\alpha_{\lambda ij}=1/\alpha_{{\rm d}ij}$.
Therefore, they allowed  $d_{ij}$
to be different
from $d_{ij}^{(0)}$ but assumed that the oscillator
frequency $\omega_{ij}$ adjusts to this new separation following
the ion-sphere rescaling, so that $\alpha_{{\omega}ij}^2=1/
\alpha_{{\rm d}ij}^3$. In contrast, we allow the separations and
oscillator frequencies to deviate independently,
and our consideration is more general.

Following Ogata {\it at al}.~\cite{oii91} (see their Eq.~(7)),
we assume that the number of
closest neighbors $j$ around the ion $i$  
in Eq.\ (\ref{pyc1}) is $\langle \nu_{ij} \rangle_{\rm av} =8\, n_j/n$
(which is appropriate for a uniform mix).
The factor `8' may be
approximate but it affects only the pre-exponent of the
reaction rate which is much less significant than the exponentially
small probability of Coulomb tunneling.
Substituting (\ref{svh2}) and (\ref{alphas}) into (\ref{pyc1}),
we obtain 
\begin{eqnarray}
   R_{ij}^{\rm pyc}&= &D_{\rm pyc}
   \,{4 n_i n_j \over 1+ \delta_{ij} }\,
   {  8 \langle Z \rangle   \over (Z_i^{1/3}+Z_j^{1/3})^3  } \,
   {S(E_{ij}^{\rm pk}) \over \hbar } 
\nonumber \\   
   && \times  { r_{{\rm B}ij} \over \widetilde{\lambda}_{ij}^{C_{\rm pl}}}\,
   \exp \left( - {C_{\rm exp} \over
   (\widetilde{\lambda}_{ij})^{1/2} }  \right).
\label{pyc2}
\end{eqnarray}

This equation has the same structure as
Eq.\ (\ref{Rth}) and can be written as
\begin{equation}
    R_{ij}^{\rm pyc}= { n_i n_j \over 1 + \delta_{ij}}
      \,S(E_{ij}^{\rm pk}) \,
     { r_{{\rm B}ij} \over \hbar }\, P_{\rm pyc} \, F_{\rm pyc},
\label{Rpyc}
\end{equation}
with
\begin{equation}
    P_{\rm pyc}=
   { 8 \langle Z \rangle  \over (Z_i^{1/3}+Z_j^{1/3})^3 } \,
   {4 D_{\rm pyc}\over {\widetilde{\lambda}_{ij}}^{C_{\rm pl}}},
   \quad
   F_{\rm pyc}=\left( - {C_{\rm exp} \over
   (\widetilde{\lambda}_{ij})^{1/2} }  \right).
\label{PFpyc}
\end{equation}

For numerical evaluations, we have
\begin{eqnarray}
  R_{ij}^{\rm pyc}& = &10^{46} \,
  C_{\rm pyc} \,
  { 8 \rho X_N x_i x_j A_i A_j \langle A \rangle Z_i^2 Z_j^2
   \over (1+ \delta_{ij}) A_c^2 } \, S(E^{\rm pk}_{ij}) 
 \nonumber \\  
   && \times 
   {\widetilde{\lambda}_{ij}}^{3-C_{\rm pl}} \,
   \exp \left( - {C_{\rm exp} \over
   (\widetilde{\lambda}_{ij})^{1/2} }  \right)~~{\rm cm}^{-3}~{\rm s}^{-1},
\label{pyc3}
\end{eqnarray}
where $C_{\rm pyc}=D_{\rm pyc}/(8 \times 11.515)$;
the density $\rho$ is
expressed in g~cm$^{-3}$ and the
astrophysical factor $S(E_{ij}^{\rm pk})$ is in MeV~barn. The reaction
energy is $E_{ij}^{\rm pk} \sim \hbar \omega_{ij}
=\alpha_{\omega ij} k_{\rm B}T_{ij}^{(p)}$.
The main parameter regulating the reaction rate
is $\widetilde{\lambda}_{ij}$ in the exponent argument. For sufficiently low
densities, $\widetilde{\lambda}_{ij}$ is very large,
strongly
suppressing the Coulomb tunneling.
With growing $\rho$ the barrier
becomes more transparent and the reaction rate increases.

For the OCP with $\alpha_\lambda=1$
Eqs.\ (\ref{pyc2})--(\ref{pyc3}) reduce to the well known equations
for zero-temperature pycnonuclear burning in
a crystalline lattice. The constants $C_{\rm pyc}$,
$C_{\rm pl}$ and $C_{\rm exp}$, obtained 
using various techniques, have been analyzed in Ref.~\cite{leandro05}.
In Table \ref{tab:pyc} we present these parameters
for three models. The first model is optimal
(seems to be the most reliable). It is the
static-lattice model of Salpeter and Van Horn \cite{svh69}
for the bcc crystal. The second and third models are phenomenological;
they have been proposed in Ref.~\cite{leandro05}, basing on the results
of Refs.\ \cite{svh69,Schramm,oii91,kitamura00}. The second model
gives the upper limit of the reaction rate and the third gives
the lower limit (for both -- bcc and fcc -- crystals).
We expect that the reaction rate in an amorphous OCP
would lie within the same limits.

Returning to an MCP, we must additionally specify the
scaling factor $\alpha_{\lambda ij}$ in Eq.\ (\ref{lambda1}).
Some attempts have been made \cite{oii91,iovh92}
to determine proper inter-ion separations $d_{ij}$
(i.e., the values of $\alpha_{{\rm d}ij}$) at $T=0$ in BIM solids
from positions of first correlation 
peaks in radial pair distribution functions
of ions, $g_{ ij}(r)$. These functions
have been calculated by Monte Carlo sampling. Such studies
require powerful computer resources and may be inconclusive
at present. This is clearly seen from similar (and simpler) attempts
of the same group \cite{oii91,iovh92,oiv93,oiiv93} to determine
deviations of inter-ion separations from the ion-sphere scaling
in BIM liquids (using first-peak positions of $g_{ij}(r)$).
The authors applied their results to study an 
effect of these deviations on thermonuclear burning in the
strong screening regime.
Our analysis of those results shows that
no statistically significant deviations from the ion-sphere
scaling have been found (and no associated
effects on nuclear burning can actually be predicted).
This conclusion is strengthened
by the critical analysis of these works
by Rosenfeld \cite{rosenfeld96,rosenfeld96b}.
Therefore, no reliable information has been
obtained on the violation of the ion-sphere scaling of inter-ion
separations in the BIM liquids and solids (on the level of a few
percent, which is most likely the real uncertainty
in the determination of $g_{ij}(r)$
peak positions in the cited publications).

Equally, proper oscillator frequencies
$\omega_{ij}$ (and parameters $\alpha_{\omega ij}$) in BIMs
could be determined from molecular dynamics simulations
but such simulations have not yet been performed.

In the absence of precise microscopic calculations
of $d_{ij}$ and $\omega_{ij}$
we naturally assume that the optimal values are
$\alpha_{{\rm d}ij}=\alpha_{\omega ij}
=\alpha_{\lambda ij}=1$. In order to maximize the reaction
rate we propose to increase $\alpha_{\lambda ij}$ (somewhat arbitrarily) by
5\%, and in order to minimize the rate we propose to reduce it by
5\%. These proposed values are also listed in Table \ref{tab:pyc}.
Notice that it is difficult to expect that
the variations of $\alpha_{{\rm d}ij}$ and $\alpha_{\omega ij}$
in Eq.~(\ref{alphas}) are fully independent. An increase
in the inter-ion separation $d_{ij}$ should cause a decrease in the
oscillator frequency $\omega_{ij}$; these variations should be
partially compensated in the factor
$\alpha_{\lambda ij}=1/(\alpha_{{\rm d}ij}^4 \alpha_{\omega ij}^2)$.

\subsubsection{Regular MCP lattice}

Now we turn to the case a regular MCP
lattice, which can be drastically different
from a uniform mix. The central point is the availability
of closest neighbors $ij$. If they are absent,
the reaction (\ref{reaction})
occurs via Coulomb tunneling of more distant $ij$ pairs
and becomes
strongly suppressed. 

The closest-neighbor condition depends on the crystal type.
For instance, consider a binary bcc crystal
composed of ions $i$ and $j$. There are eight pairs of closest neighbors in a
basic cubic cell
(formed by one ion in the center of the cell and any
other ion in a vertex). If all ions are of the same type
(e.g., $x_i=1$),
then all eight pairs participate in the same reaction
$ii$
(the OCP case).
In the BIM case ($x_i=x_j=\frac12$) we have an ion $i$
in the center of the cell and
ions $j$ in vertices, and all
eight pairs of closest neighbors participate
only in the reaction $ij$. The reactions $ii$ and $jj$ will
be strongly blocked because any pairs $ii$ and $jj$
are not closest neighbors. Then the ion $i$ in the center of
the basic cell will be able to react with six ions $i$
in centers of adjacent cells. The equilibrium distance between
these pairs is a factor of $2/\sqrt{3}\approx 1.155$ larger,
than between the closest neighbors, which
will exponentially suppress the $ii$ reaction rate. 

\begin{figure}[b]
\begin{center}
\epsfysize=80mm
\epsffile[20 160 475 590]{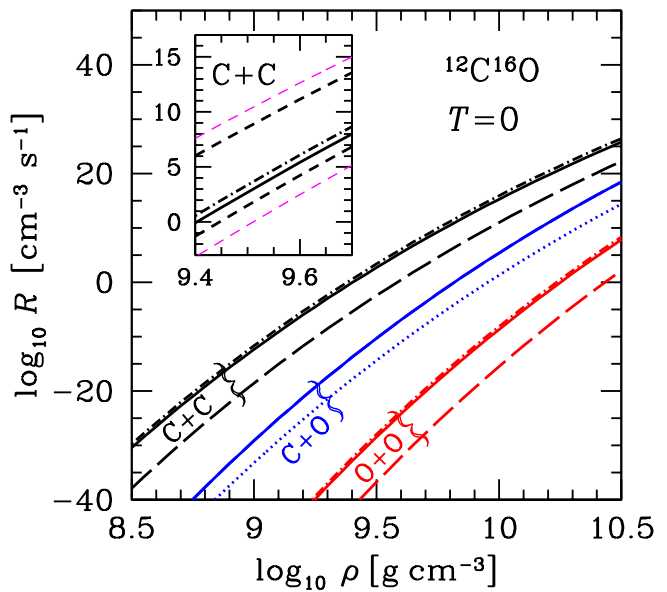}
\caption{(color online)  Rates of pycnonuclear C+C, C+O and O+O reactions
versus density for the optimal burning model. The dot-and-dashed
lines show C+C burning in a pure carbon crystal and O+O burning
in a pure oxygen crystal. Other lines are for the CO mixture
($x_{\rm C}=0.5$); the solid lines refer to a uniform BIM;
the long-dashed lines are for a regular CO bcc crystal; the dots are
for phase-separated matter. The inset
shows the same curves for the C+C reaction on a larger scale;
the thicker short-dashed lines  give the maximum and
minimum reaction rates for the uniform CO mixture in the ion-sphere
model ($\alpha_\lambda=1$); the thin short-dashed lines are the
same but allowing for the variation of $\alpha_\lambda$
(see the text for details).
}
\label{fig:pyc}
\end{center}
\end{figure}

One can construct more complicated MCP
lattice structures and formulate appropriate blocking conditions.
The strongest blocking of pycnonuclear reactions
is expected in a regular crystal with many components.
In that case the probability to find specified closed neighbors
$ij$ could be very selective.

\subsubsection{Other MCP structures}

If the matter consists of domains of separated phases,
pycnonuclear burning occurs mainly within these domains.
The rate of the reaction $ii$
in domains containing ions $i$ can be calculated from
Eq.\ (\ref{pyc3}), assuming OCP ($x_i=1$),
and then diluted by a volume fraction occupied by the phase $i$. The burning
of different ions $i$ and $j$ occurs on interfaces between
corresponding domains. Roughly, in a BIM with
spherical bubbles containing $N$ less abundant
ions ($i$ or $j$) the OCP reaction rate $ii$ can be diluted with
respect to the rate in a uniformly mixed state by a
factor of $1/N^{2/3}$ (because the reactions occur on interfaces).

Pycnonuclear burning in a lattice
can be drastically
affected by lattice impurities and imperfections
(see, e.g., Ref.\ \cite{svh69}). We expect that the effect of
impurities or imperfections can be included
in Eq.\ (\ref{pyc3}) if we treat them as members of the MCP.
Because the burning is mainly regulated by the parameter
$\widetilde{\lambda}_{ij}$, the rate should be extremely sensitive to
variations of equilibrium distances and oscillation frequencies
for the reacting nuclei (see Eq.\ (\ref{lambda1})).
Small variations can induce exponentially huge jumps or
drops of the Coulomb tunneling probability (for instance, in response
to the decrease or increase of inter-ion separations).
The impurities and imperfections may be rare but give
the leading contribution into the reaction rate.

Moreover, different pairs $ij$ in an MCP
may be exposed to different local conditions and have
different separations and oscillator frequencies. One can 
incorporate these effects by introducing the
averaging $\langle \ldots \rangle_{\rm av}$ over an ensemble of such
pairs in Eq.\ (\ref{pyc1}).

For illustration, in Fig.\ \ref{fig:pyc} we show
the density dependence of C+C, C+O and O+O pycnonuclear
reactions in carbon-oxygen BIMs.
The astrophysical factors are taken from Section \ref{nuclear}.
In the main part of the figure we use the optimal reaction model
from Table~\ref{tab:pyc}. The dot-and-dashed lines show
the C+C burning in a pure carbon matter ($x_{\rm C}=1$)
and O+O burning in a pure oxygen matter ($x_{\rm C}=0$) while other
lines are for the CO mixture with $x_{\rm C}=\frac12$.
The solid lines present the reaction rates in the uniform CO mixture.
For the C+C and O+O reactions these lines go slightly lower than
the dash-and-dot lines because of the reduced amount of
carbon and oxygen in a BIM (as compared to pure carbon
or oxygen matter). The long-dashed lines give the reaction
rates in the bcc CO regular crystal. 
The C+O burning in this crystal has the same rate as in the
CO uniform mix (the solid line), in our model.
In order to illustrate the blocking
effect (produced by the absence of 
closest CC or OO neighbors in the crystal)
we have calculated the C+C and O+O reaction rates
using Eq.~(\ref{pyc3}) but increasing the distances between
the reacting nuclei in Eq.~(\ref{alphas}) by a factor of
$\alpha_{{\rm d}ii}=2/\sqrt{3}$, where $i$=C or O (see above).
Because the local oscillator frequencies for these CC and OO
pairs are unknown, we have assumed the ion-sphere
rescaling ($\alpha_{\omega ii}^2=1/\alpha_{{\rm d}ii}^3$).
The blocking effect is strong, reducing the reaction
rates by 5--7 orders of magnitude. Nevertheless, even with this
blocking, the C+C reaction is much faster than the C+O and O+O ones
because of much lower Coulomb barrier. Finally, we have also
calculated the BIM reaction rates assuming phase separation of C and O.
We do not plot the respective C+C and O+O reaction rates
because they are almost indistinguishable from the corresponding
dot-and-dashed lines
(the rates are only twice lower than
in the respective OCPs). However, the C+O burning in this case
can be strongly suppressed because it occurs only at interfaces
between separated phases. For example, the dotted line shows
such a C+O burning rate assuming separation into domains which contain
$10^6$ ions.

The inset in Fig.~\ref{fig:pyc} displays the C+C 
reaction rate on a larger scale. The solid and
dot-and-dashed lines are the same as in the main part of the figure.
Short dashed lines present the maximum and minimum reaction rates
for the uniform CO mixture; they reflect
uncertainties of existing theoretical models for pycnonuclear
burning. These lines
are calculated using the parameters from Table \ref{tab:pyc}.
Thicker short-dashed lines give the maximum and minimum reaction
rates using the ion-sphere model (neglecting deviations
from the ion-sphere rescaling;
i.e., assuming $\alpha_{\lambda \rm CC}=1$).
Thinner short-dashed lines are the estimated maximum and
minimum reaction rates taking into account possible deviations
from the ion-sphere model. The current theoretical uncertainties
of the reaction rate are really large.

\begin{table*}
\caption[]{Coefficients in
the interpolation expressions for a reaction rate
for the
optimal model of nuclear burning and for the models
which maximize and minimize the rate.
The parameters $C_T$, $\alpha_{\lambda ij}$, $\alpha_{{\omega}ij}$ are
different for MCP and OCP (the values
for OCP \cite{leandro05} are given in brackets).
For an MCP, the models assume a uniformly mixed state (see the text
for details).}
\label{tab:pyc}
\begin{center}
\begin{tabular}{lccccccc}
\hline
\hline
~Model~& ~~~$C_{\rm exp}$~~~  & ~~$C_{\rm pyc}$~~  &~~ $C_{\rm pl}$~~~ &
~~~~$C_T$~~~~~ &~~~~~$\alpha_{\lambda ij}$~~~~~&~~~~~$\alpha_{\omega ij}$
~~~~~&~~~$\Lambda$~~~    \\
\hline
Optimal & 2.638 & 3.90  & 1.25  & 0.724 (0.724)   & 1~~~ (1) &  1~~~ (1) & 0.5   \\
Maximum rate &2.450 & 50  &  1.25  & 0.840 (0.904) & 1.05 (1) & 0.95 (1)  & 0.35 \\
Minimum rate &2.650 & 0.5  &  1.25  & 0.768 (0.711)& 0.95 (1) & 1.05 (1)  & 0.65 \\
\hline
\hline
\end{tabular}
\end{center}
\end{table*}

\subsection{Thermally enhanced pycnonuclear regime}
\label{thermopyc}

This regime operates in the temperature range
$0.5\,T_{ij}^{(p)} /\ln ( T_{ij}^{(l)}/T_{ij}^{(p)} ) \lesssim T
\lesssim 0.5\,T_{ij}^{(p)}$ (see Ref.\ \cite{svh69}). 
At these temperatures, the majority
of nuclei occupy ground states in their potential wells
but the main contribution into the reaction rate comes
from a small amount of nuclei which occupy excited bound
states. This regime is difficult for
theoretical studies. We will follow the approach
of Ref.~\cite{leandro05}, which is based on
an analytical approximation of the WKB calculations performed
by Salpeter and Van Horn~\cite{svh69}. We generalize this approach
to an MCP in the same manner as in
Section \ref{pycnozero}, by  rescaling $\lambda$
and $T_p$ in accordance with Eqs.\ (\ref{lambda}) and (\ref{notations}).
In this case 
\begin{equation}
    { R_{ij}^{\rm pyc}(T) \over R_{ij}^{\rm pyc}(0)} -1 =
     {\Omega \over  \widetilde{\lambda}_{ij}^{1/2} } \,
     \exp
     \left(
         - \Lambda \,{\widetilde{T}_{ij}^{(p)}  \over T} +
         { \Omega_1 \over (\widetilde{\lambda}_{ij})^{1/2}}\,
        {\rm e}^{ - \Lambda \widetilde{T}_{ij}^{(p)} / T}
     \right),
\label{pyctherm}
\end{equation}
where $\widetilde{T}_{ij}^{(p)} \equiv \hbar \omega_{ij}/ k_{\rm B}
=\alpha_{\omega ij} T_{ij}^{(p)}$,
while $\Omega$, $\Omega_1$, and $\Lambda$ are model-dependent
dimensionless constants.
We adopt $\alpha_{\omega ij}=0.95$ to maximize the reaction rate
and $\alpha_{\omega ij}=1.05$ to minimize it (Table \ref{tab:pyc}).
In analogy with Ref.\ \cite{leandro05}
the characteristic energy of the reacting nuclei
can be taken in the form
\begin{equation}
  E_{ij}^{\rm pk} \approx  \hbar  \omega_{ij} +
   \, { Z_i Z_j e^2 \over a_{ij}} \,
  \exp \left(-\Lambda \,{ \widetilde{T}_{ij}^{(p)} \over T }\right).
\label{pk-pyctherm}
\end{equation}
The first term is the reaction energy in the
zero-temperature pycnonuclear regime,
while the second term describes 
an increase of $E_{ij}^{\rm pk}$ with growing temperature
in the thermally enhanced pycnonuclear regime.
Equation (\ref{pk-pyctherm}) is approximate 
(based on the results of Ref.~\cite{svh69} as explained in \cite{leandro05}) 
but we expect that it correctly
reflects the main features of the accurate expression
(to be derived in future, more elaborated
calculations). The analogous expression (26) for OCP in
Ref.\ \cite{leandro05} contains two new free factors,
$C_1$ and $C_2$, in the first and second terms, respectively.
They were set $C_1=C_2=1$ in numerical calculations \cite{leandro05}
because the theory \cite{svh69} is insufficiently precise
to determine them.
We do not introduce these factors here to avoid
additional unknowns, but they could be introduced
in the future. An uncertainty in $E_{ij}^{\rm pk}$
should not greatly affect the reaction rates. 

After Salpeter and Van Horn \cite{svh69}
the thermally enhanced pycnonuclear burning in an OCP was studied
by Kitamura and Ichimaru \cite{ki95}
assuming that the reacting nuclei move in an angle-averaged,
radial static mean-field potential determined from
Monte Carlo sampling of classical Coulomb systems.
Although this approach is less justified than
the WKB approximation of Ref.\ \cite{svh69}, the results
are in a reasonable agreement (see Ref.~\cite{leandro05}
for details).  Kitamura \cite{kitamura00} generalized
the results of Ref.~\cite{ki95} to the case of BIMs
using the ion-sphere rescaling rule
($\alpha_{\lambda ij}=\alpha_{\omega ij}=1$).

The thermally enhanced pycnonuclear burning  
is as sensitive to the microphysical structure of
the MCP as the zero-temperature pycnonuclear burning
(Section \ref{pycnozero}). The above comments refer to
a uniformly mixed MCP. In the case of a regular MCP lattice the reaction
will be suppressed by the same blocking effects as
discussed in Section \ref{pycnozero}. The case of phase separation
has the same features as at $T=0$. Illustrative examples will
be given in Section \ref{ignition}.

\subsection{The intermediate thermo-pycnonuclear regime}
\label{intermed}

This regime is realized at
temperatures $T_{ij}^{(p)}/2 \lesssim T \lesssim
T_{ij}^{(p)}$ which separate the domains of classical and
quantum motion of the reacting nuclei.
The calculation of the reaction rate in
this regime is most complicated. We will describe this rate by a
phenomenological expression presented below.

The reaction is mainly determined by
the nuclei which become slightly unbound and can
move freely through the dense matter, fusing not only with the closest
neighbors (pycnonuclear regime), but with other nuclei
(thermonuclear regime).
We expect that the transition
from the pycnonuclear to the thermonuclear regime with the
growth of temperature in a uniformly mixed
MCP is sufficiently smooth.
When the number of freely reacting 
nuclei becomes large, the dependence of the reaction rate
on the details of the MCP microstructure should disappear.

\subsection{Single analytical approximation in all regimes}
\label{together}

Our phenomenological expression for the 
temperature and density dependent reaction rate, which
combines all the five burning regimes
and {\it assumes a uniformly mixed MCP at low temperatures},
is a straitforward generalization
of the expression for the OCP
considered in Ref.~\cite{leandro05},
\begin{eqnarray}
&&    R_{ij}(\rho,T) =  R_{ij}^{\rm pyc}(\rho) + \Delta R_{ij}(\rho,T), 
\nonumber \\
&&    \Delta R_{ij}(\rho,T) =  \frac{n_in_j}{1 + \delta_{ij}} \,
    {S(E_{ij}^{\rm pk}) \over \hbar} \,
     r_{{\rm B}ij}\, P \, F,
\nonumber \\
&&     F = \exp \left(-\widetilde{\tau}_{ij}
     +C_{\rm sc}\widetilde{\Gamma}_{ij}\,
     \varphi \,{\rm e}^{-\Lambda \widetilde{T}_{ij}^{(p)}/T}
     -\Lambda\, {\widetilde{T}_{ij}^{(p)} \over T} \right),
\nonumber \\
&&     P= { 8 \, \pi^{1/3} \over \sqrt{3}\; 2^{1/3}} \,
       \left(E_{ij}^{\rm a} \over k_{\rm B} \widetilde{T} \right)^\gamma.
\label{overall}
\end{eqnarray}
In this case, $\varphi = \sqrt{\Gamma_{ij}}/
[(C^{\rm sc}_{ij}/\zeta_{ij})^4+\Gamma^2_{ij}]^{1/4}$;
$R_{ij}^{\rm pyc}(\rho)$ is the density dependent pycnonuclear reaction
rate at zero temperature 
discussed in Section \ref{pycnozero}; $\Delta R_{ij}(\rho,T)$
is the density and temperature dependent part including an
exponential function $F$ and a pre-exponent $P$. The quantities
$\widetilde{\tau}_{ij}$ and  $\widetilde{\Gamma}_{ij}$ are
similar to the familiar quantities $\tau_{ij}$ and $\Gamma_{ij}$, but
contain a ``renormalized'' temperature $\widetilde{T}$,
\begin{eqnarray}
  &&  \widetilde{\tau}_{ij}=3\, \left(\pi \over 2 \right)^{2/3}
    \left(E_{ij}^{\rm a} \over
     k_{\rm B} \widetilde{T} \right)^{1/3},
\nonumber \\
  &&  \widetilde{\Gamma}_{ij} = {Z_iZ_j e^2 \over
    a_{ij} k_{\rm B} \widetilde{T} },
    \quad
    \widetilde{T} = \sqrt{ T^2+C_T^2(T_{ij}^{(p)})^2},
\label{renorm}
\end{eqnarray}
where $C_T$ is a dimensionless
renormalization parameter specified below. 
Equations (\ref{overall}) and (\ref{renorm})
are analogous to Eqs.~(27) and (28) of Ref.~\cite{leandro05}. 
The term $\Delta R_{ij}(\rho,T)$ and the renormalized 
temperature $\widetilde{T}$ are introduced to match
the equations in thermonuclear and pycnonuclear regimes.
The renormalized temperature reflects the fact
that the thermal energy $k_{\rm B} T$ of plasma
ions in the thermonuclear case is replaced by
a temperature-independent zero-point energy
in the pycnonuclear case.

For high temperatures $T \gg {T}_{ij}^{(p)}$ 
we have $\widetilde{\tau}_{ij} \to \tau_{ij}$,
$\widetilde{\Gamma}_{ij}
\to \Gamma_{ij}$, and $\widetilde{T} \to T$. In this case
$\Delta R_{ij}(\rho,T) \to R_{ij}^{\rm th}(\rho,T)
\gg R_{ij}^{\rm pyc}(\rho)$, and Eq.\ (\ref{overall}) reproduces the
thermonuclear reaction rate (Sections\ \ref{thermo} and
\ref{thermoscreen}). At low temperatures $T \lesssim T_{ij}^{(p)}$ the
quantities $\widetilde{\tau}_{ij}$, $\widetilde{\Gamma}_{ij}$ and
$\widetilde{T}$ contain ``the quantum
temperature'' $T_{ij}^{(p)}$,
determined by zero-point ion vibrations, 
rather than the real temperature $T$.
In the limit of $T \to 0$ we obtain
$\widetilde{\Gamma}_{ij}=1/[(\lambda_{ij})^{1/2}\,(72 \pi)^{1/6} \, C_T]$
and 
$
\widetilde{\tau}_{ij}=
3 \, (\pi/\lambda_{ij})^{1/2} \, 
/
(2^{7/6} \,C_T^{1/3} \, ).$

Following Ref.~\cite{leandro05}
we require that at $T \ll T_{ij}^{(p)}$ the factor
$\exp(-\widetilde{\tau}_{ij})$ in the exponential
function $F$, Eq.\ (\ref{overall}), reduces to
$\exp(-C_{\rm exp}/(\widetilde\lambda_{ij})^{1/2})$.
This would allow us to obey
Eq.\ (\ref{pyctherm}) by satisfying the equality
\begin{equation}
   {3\, \sqrt{\pi} /( 2^{7/6}\, C_T^{1/3})}
   =C_{\rm exp}\, (\alpha_{\lambda ij})^{-1/2}.
\label{Crelat}
\end{equation}
The double exponent factor in $F$, Eq.\ (\ref{overall}),
will correspond to the
double exponent factor in  Eq.\ (\ref{pyctherm}).
Taking $C_{\rm exp}$ and
$\alpha_{\lambda ij}$ from Table \ref{tab:pyc} we can determine $C_T$.
These parameters are also listed in Table \ref{tab:pyc}.
In the MCP they are different from those
in the OCP because in the MCP we introduce
an additional parameter $\alpha_{\lambda ij}$
(Section \ref{pycnozero}).
The values of $C_T$ for the OCP \cite{leandro05}
are given in Table \ref{tab:pyc} in brackets.

Finally, the quantity $\gamma$ in Eq.\ (\ref{overall})
and the reaction energy $E_{ij}^{\rm pk}$ in
the astrophysical factor $S(E_{ij}^{\rm pk})$
can be chosen in the same way as in Ref.\ \cite{leandro05},
\begin{equation}
    \gamma=\left(T^2 \gamma_1 +
    (\widetilde{T}_{ij}^{(p)})^2 \gamma_2 \right)/
    \left(T^2+(\widetilde{T}_{ij}^{(p)})^2 \right).
\label{gamma}
\end{equation}
\begin{equation}
    E_{ij}^{\rm pk}= \hbar \widetilde{\omega}_{ij}^{(p)}
    + \left( {Z_iZ_j e^2 \over a_{ij}} +
    {k_{\rm B}T \tau_{ij} \over 3} \right) \,
    \exp \left( - { \Lambda \, \widetilde{T}_{ij}^{(p)} \over T} \right),
\label{peak}
\end{equation}
where $\gamma_1=2/3$ and $\gamma_2=(2/3)\,(C_{\rm pl}+0.5)$.

Thus, we propose to use the analytic expression (\ref{overall})
for the reaction rate in a uniformly mixed 
MCP with the following parameters:\\
    (1) The parameter
$C_{\rm sc}$ of strongly screened thermonuclear burning 
is given by Eq.~(\ref{salpeter}).\\
    (2) The parameters
$C_{\rm exp}$, $C_{\rm pyc}$, $C_{\rm pl}$,
$\alpha_{\lambda ij}$
of zero-temperature pycnonuclear burning,
and the parameters $\alpha_{{\omega}ij}$,
$\Lambda$ and $C_T$ of
thermally enhanced pycnonuclear burning are given in Table \ref{tab:pyc}.\\

In this way we obtain (Table \ref{tab:pyc}) three models 
for any given non-resonant nuclear
fusion reaction (\ref{reaction}) in a  
uniformly mixed MCP. One is the optimal
model, the second gives the maximum reaction rate, and the third
the minimum reaction rate.
For the OCP, it is sufficient to
set $\alpha_{\lambda  ij}=\alpha_{\omega ij}=1$, which
reduces the present results to the results of Ref.~\cite{leandro05}. 

The uncertainties of the reaction rate
become larger if a cold MCP  
forms a regular lattice or undergoes a phase separation
or contains impurities and defects. 
All these cases 
can be approximately taken into account in the same 
way as discussed in Section \ref{pycnozero}.
For instance, a reaction in a regular lattice 
can be strongly blocked by the absence
of closest reacting neighbors (Sections \ref{pycnozero} and
\ref{thermopyc}; also see Section \ref{ignition}). 

Our formula for a uniformly mixed MCP gives a smooth
behavior of the reaction rate as a function of
temperature and density, without jumps at the solidification
point (in analogy with an OCP, see Ref.\ \cite{leandro05}).
In the cases of other MCP microstructures such jumps may appear.

Our formula is flexible. Its parameters could be
tuned when new microscopic calculations
of reaction rates appear in the future. Moreover, the formula can be
improved even if a new information on MCP properties
(not on reaction rates directly) appear in the
literature (for instance, on the deviations from the ion-sphere
scaling at $T=0$).

More complicated expressions for the reaction rates
in the OCP and uniform BIMs were proposed
by Kitamura \cite{kitamura00}.
His expressions are mainly
based on the results of Refs.\ \cite{oii91,iovh92,oiv93,ki95,ogata97}
(in the different regimes)
which are not free of approximations
(see Ref.~\cite{leandro05} for details).
His expressions for BIMs
are obtained assuming the ion-sphere rescaling
rule ($\alpha_{\lambda ij}=\alpha_{\omega ij}=1$)
and are, therefore, more restricted than our expression.
Their derivation implies that they are valid for uniformly mixed BIMs.
In particular, they do not take into account blocking effects
in regular binary lattices.

In contrast to our formula,
Kitamura took into account the effects of electron screening
(finite polarizability of the electron gas).
However, these effects are relatively
weak; their strict inclusion in the pycnonuclear regime
is complicated.
We do not include them but, instead, take into account
theoretical uncertainties of the reaction rates without
electron screening. The results
of Kitamura \cite{kitamura00} for an OCP
lie well within these uncertainties. His results for BIMs
in the thermonuclear regime with strong screening
and essentially different charges of reacting nuclei
are less accurate than our results
(Section \ref{thermoscreen}).

\section{Nuclear burning in a carbon-oxygen mixture}
\label{ignition}

\begin{figure}[tbh]
\begin{center}
\epsfysize=80mm
\epsffile[20 160 475 590]{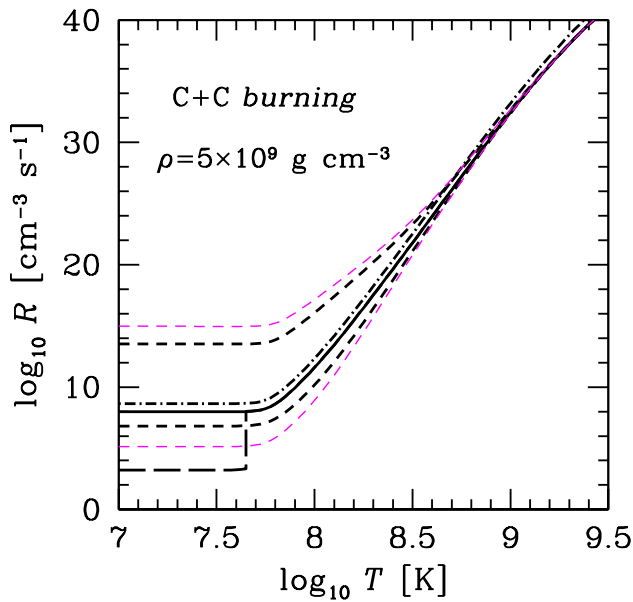}
\caption{(color online) The C+C reaction rate
versus temperature at $\rho= 5 \times 10^9$ g~cm$^{-3}$.
The dot-and-dashed line is the optimal model
for a pure carbon matter;
other lines are for a CO mixture
with $x_{\rm C}=0.5$. The solid line is the optimal model
for a uniform mixture; the long-dashed line is for the regular bcc
CO crystal at low temperatures (the vertical part indicates the
melting point).
Thicker short-dashed lines give the maximum and
minimum reaction rates for the uniform CO mixture in the ion-sphere
model ($\alpha_\lambda=1$); thinner short dashed-lines are the
same but allowing for deviations from the ion-sphere model
(see the text for details).
}
\label{fig:temper}
\end{center}
\end{figure}

For illustration of our results,
we analyze nuclear reactions in a dense $^{12}$C\,$^{16}$O
mixture. Figure \ref{fig:temper} shows the temperature
dependence of the C+C reaction rate
at $\rho=5 \times 10^9$ g~cm$^{-3}$ (for the same
reaction models as in Fig.\ \ref{fig:pyc}).
The dot-and-dashed line is the optimal model
(Table \ref{tab:pyc}) for a pure carbon matter
(from Ref.\ \cite{leandro05}). Other lines are for
CO mixtures
with equal numbers of C and O nuclei ($x_{\rm C}=\frac12$).
The solid line is the optimal model for the uniform mixture;
the thicker short-dashed lines show the maximum and minimum reaction
rates in such a mixture in the ion-sphere approximation
($\alpha_{ \lambda \rm CC}=\alpha_{\rm \omega CC}=1$);
the thinner short-dashed lines are the same but beyond the
ion-sphere approximation (note that
variations of $\alpha_{\rm \omega CC}$
appear to be much less important than variations
of $\alpha_{\rm \lambda CC}$). The long-dashed line is for
a CO regular lattice (in the same approximation which has been
used in Fig.\ \ref{fig:pyc}). A sharp jump
of this curve is associated with the melting of the crystal
(Fig.\ \ref{fig:diag}), which destroys the blocking
of the C+C burning and amplifies the reaction rate.

Figure \ref{fig:temper} shows the reaction rates in all
burning regimes (except for the classical thermonuclear burning
which would require higher temperatures; see Fig.\ \ref{fig:diag}).
The horizontal parts of the curves for $\log T{\rm [K]} \lesssim 7.7$
refer to the zero-temperature pycnonuclear burning;
the respective reaction rates are independent of $T$ as discussed
in Section \ref{pycnozero} and displayed in Fig.~\ref{fig:pyc}.
The temperature range $7.7 \lesssim \log T{\rm [K]} \lesssim 8.3$
corresponds to the thermally enhanced pycnonuclear regime.
The reaction rate starts to grow up with increasing $T$
(Section \ref{thermopyc}).
The rate 
remains highly uncertain
because of the same reasons as in the zero-temperature
pycnonuclear regime.
The next temperature
range $8.3 \lesssim \log T{\rm [K]} \lesssim 8.6$ corresponds
to the intermediate thermo-pycnonuclear burning
(Section \ref{thermopyc}). Theoretical
uncertainties of the reaction rate become smaller.
Finally, the temperature range $ \log T{\rm [K]} \gtrsim 8.6$
refers to the thermonuclear burning with strong plasma
screening. The theoretical uncertainties 
become much smaller although the enhancement of the reaction
rate by the plasma screening effects is huge;
with increasing $T$ this enhancement weakens and the
reaction rate matches the classical thermonuclear rate
(see Fig.~6 of Ref.\ \cite{leandro05}). The presence of oxygen
slightly reduces the C+C reaction rate (by reducing the amount
of carbon nuclei at a given density).

\begin{figure}[t]
\begin{center}
\epsfysize=80mm
\epsffile[20 160 475 590]{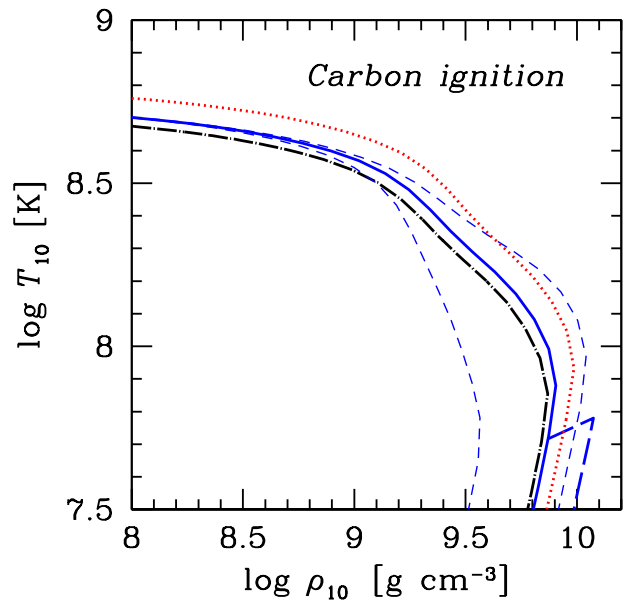}
\caption{(color online) Carbon ignition curves in $^{12}$C\,$^{16}$O matter.
The dot-and-dashed line is the optimal model for carbon burning
in pure carbon matter. The solid and dotted lines are optimal
models for uniform CO mixtures with $x_{\rm C}=$0.5 and 0.1,
respectively. Other lines are for CO BIMs with $x_{\rm C}=0.5$.
The short-dashed lines give the highest and the lowest
theoretical ignition curves for uniform mixtures. The long-dashed
line is for the CO bcc crystal at low temperatures.
}
\label{fig:igni}
\end{center}
\end{figure}

Based on our expression for the reaction rates, in Fig.~\ref{fig:diag} 
we plot the $T- \rho$ domains (shaded strips), where the
C+C, C+O and O+O reactions are most important.
The domains for the C+C and O+O reactions are presented for pure carbon
or pure oxygen matter
($x_{\rm C}=1$ and $x_{\rm C}=0$, respectively). For the C+O reaction
we have taken the CO mixture with $x_{\rm C}=\frac12$. A domain for
any reaction $ij$ is restricted by two lines, along which the
characteristic burning time $\tau_i=n_i/R_{ij}$ of nuclei $i$
is constant (taken to be $10^6$ years for a lower line and
1~year for an upper line, for example).
Above the upper line the reaction
$ij$ is so fast, that the nuclei $i$ cannot survive for a long time.
Below the lower line the reaction is so slow that the nuclei $i$
survive almost forever. Therefore,
the strips represent the temperature and density domains
of greatest relevance
for the carbon and oxygen nucleosynthesis
through the reactions under discussion.
For determining these domains, we have taken the optimal model
from Table \ref{tab:pyc}. The domains do not change significantly
under variations of fractional numbers of C and O within reasonable
limits. For densities
$\rho \lesssim 10^9$ g~cm$^{-3}$
the strips are almost horizontal; nuclear burning in them proceeds
in the thermonuclear regime and the reaction rates depend
mainly on the temperature. In contrast, the strips become almost vertical
at low temperatures reflecting the pycnonuclear burning regime
where the reaction rates depend mainly on the density.

The strips show a strong heterogeneity of the different
reactions. It is evidently caused by different heights of Coulomb barriers.
With increasing $\rho$ and/or $T$ in the CO matter, carbon will
burn first in the C+C reaction and could be burnt almost completely
before reaching the $T-\rho$ domain, where the C+O reaction can be
efficient. 

Finally, we have studied the carbon ignition curve, which is
a necessary ingredient
for modeling nuclear explosions of massive white dwarfs
(supernova Ia events)
and carbon explosions in accreting
neutron stars (superbursts).
The ignition curve is usually determined as the line in the
$T-\rho$ plane (Fig.\ \ref{fig:igni}), where
the nuclear energy generation rate equals the local
neutrino energy losses.  At higher $T$ and $\rho$
(above the curve) the nuclear energy generation exceeds the
neutrino losses (which cool the matter)
and carbon ignites. We have calculated such curves
for CO mixtures. All reactions
(C+C, C+O, and O+O) have been taken into account but the C+O and O+O reactions
have appeared to be unimportant owing to
the heterogeneity of nuclear burning. The presence of oxygen
affects carbon ignition only through the C+C reaction rate and
the neutrino emission rate.
The neutrino energy losses have been assumed
to be produced by plasmon decay and by electron-nucleus
bremsstrahlung. The neutrino emissivity owing to plasmon decay
has been obtained from extended tables calculated by M.~E.~Gusakov
(unpublished); they are in good agreement with the results of Itoh
{\it et al}.\ \cite{itohetal92}. The neutrino bremsstrahlung emissivity
has been calculated using the formalism of Kaminker {\it et al}.\
\cite{kaminkeretal99}, which takes into account electron band
structure effects in crystalline matter. For a CO mixture,
this neutrino emissivity has been determined using the linear
mixture rule.

The dot-and-dashed line
in Fig.\ \ref{fig:igni} shows
the carbon ignition curve,
calculated using the optimal model of carbon burning
in a pure carbon matter ($x_{\rm C}=1$; from Ref.\ \cite{leandro05}).
The solid and dotted lines are the same curves in CO mixtures
with $x_{\rm C}=0.5$ and $0.1$, respectively
(assuming the optimal reaction model and a uniform mixture
at low temperatures).
At $\rho \lesssim 10^9$~g~cm$^{-3}$ the curves depend weakly
on the density because carbon burns in the thermonuclear regime.
At $T \lesssim 10^8$~K the curves depend weakly on the temperature
because carbon burns in the pycnonuclear regime. A strong bending of the
curves in the density range from $\sim 10^9$~g~cm$^{-3}$
to $\sim 3 \times 10^9$~g~cm$^{-3}$ is associated with the
transition from the thermonuclear
burning to the pycnonuclear one.
With decreasing the carbon
fraction, the ignition curve shifts to higher $T$ and $\rho$,
mainly because of the decrease of the C+C reaction rate.

The short-dashed lines in Figure \ref{fig:igni}
show the uncertainty of the solid ignition curve
($x_{\rm C}=\frac12$, a uniform CO mixture) associated with
the uncertainties of the reaction rates (assuming the maximum and
minimum reaction rates from Table \ref{tab:pyc}).
In the thermonuclear regime the uncertainties are
small, while in the pycnonuclear regime they are substantial.
The long-dashed line shows the ignition curve calculated
under the assumption that a regular bcc CO lattice is formed
in the CO mixture ($x_{\rm C}=\frac12$) after the crystallization.
The blocking of the C+C reaction rate by oxygen ions in the
bcc lattice shifts the ignition curve to higher $\rho$.
The sudden break of this line is associated with the
crystallization (analogous to the break
of the long-dashed line in Fig.\ \ref{fig:temper}).

The carbon ignition curve
obtained by equating the nuclear energy generation 
and the neutrino losses becomes unreliable for
$T \lesssim 10^8$~K (e.g., Ref.\ \cite{leandro05}).
The main reason is that this curve falls in the $T-\rho $ domain,
where the characteristic carbon burning time is unrealistically
large (exceeds the Universe age). In addition, the neutrino emission
becomes very slow, inefficient for carrying away the
nuclear energy;
thermal conduction can be much more efficient.
As a result, the carbon ignition condition becomes nonlocal,
complicated, and dependent on a specific model
(a neutron star or a white dwarf, etc.).

\section{Conclusions}
\label{conclusions}

We have studied the problem of Coulomb barrier penetration
for non-resonant nuclear fusion reactions in a dense MCP
of atomic nuclei. We have considered all
five nuclear burning regimes (Sections \ref{thermo}--\ref{intermed})
and analyzed calculations of nuclear reaction rates in an MCP
for these regimes, available in the literature.
We have proposed (Section \ref{together}) a unified phenomenological
expression for the reaction rate valid for all regimes.
It generalizes an analogous expression proposed recently \cite{leandro05}
for an OCP. The expression contains
several parameters which can be varied
to account for current theoretical uncertainties of the reaction
rates.

Our main conclusions are:
\begin{enumerate}

\item The reaction rates in the thermonuclear regimes
(with weak and strong plasma screening) can be calculated
sufficiently accurately. In the regime of strong screening
and for reacting nuclei with non-equal charges, our expression
is more accurate than those proposed in the literature
(Section \ref{thermoscreen}).

\item The reaction rates in other regimes (zero-temperature and
thermally enhanced pycnonuclear regimes; intermediate thermo-pycno
nuclear regime) are much less certain. They are
very sensitive to currently unknown
microphysical correlation properties in
an MCP (a uniform mix,
a regular crystalline lattice, a phase separated matter,
a matter with impurities and defects); they are much richer
in physics than in the OCP case.  

\item At low temperatures, we have mainly considered reactions
in a uniform mix. Other MCP microstructures 
can strongly decrease or increase the reaction rates.
For instance, the reactions in a regular 
MCP lattice can be strongly suppressed
by the absence of nearby reacting nuclei.

\item Our phenomenological expression can be improved
(Section \ref{together})
after new calculations of the reaction rates or
main properties of the MCP are performed. It would be important to
know the actual microstructure of the MCP at low temperatures
(first of all, the availability of closest neighbors,
local separations and oscillation frequencies
of neighboring nuclei, particularly in the presence
of impurities and lattice defects).

\item Although our main formula in Section \ref{together}
assumes a uniform mix at low $T$, the presented results are sufficient
to understand qualitatively the reaction rates for other
cases (following prescriptions of Section \ref{pycnozero}).

\end{enumerate}

For illustration, we have considered (Section \ref{ignition})
C+C, C+O, and O+O nuclear reactions in a dense carbon-oxygen mixture,
that is important for the structure and evolution of
massive white dwarfs (supernova Ia explosions)
and accreting neutron stars (as sources of superbursts).
For this purpose we have calculated and parameterized
the appropriate astrophysical factors (Section \ref{nuclear}).
The main results of our analysis are as follows:

\begin{enumerate}
\item
The ranges of densities and temperatures,
where C+C, C+O, and O+O reactions are most important,
look like narrow regions in the temperature-density diagram
(Fig.~\ref{fig:diag}); the regions do not strongly overlap
which means strong heterogeneity of these
reactions.

\item

With increasing density and/or temperature, carbon 
starts burning first in the C+C reaction (because carbon
nuclei have lower Coulomb barrier);
this reaction is most important for the nuclear
evolution of CO mixtures.

\item

Carbon burning in the C+C reaction is affected by the
presence of oxygen.
The effect is simple
in the thermonuclear regimes but more complicated
in other regimes (at low temperatures).

\item

Carbon ignition in a CO mixture occurs (Fig.\ \ref{fig:igni})
in thermonuclear regimes as long as
$\rho \lesssim 10^9$ g~cm$^{-3}$ (and $T \gtrsim 3 \times 10^8$~K).
It can be calculated
quite accurately.
With decreasing the fraction
of carbon, the ignition curve shifts to higher $\rho$ and $T$.

\item
At $\rho \gtrsim 10^9$ g~cm$^{-3}$ and $T \lesssim 3 \times 10^8$~K
the ignition condition becomes uncertain
(Section \ref{ignition}). The formation
of a regular CO lattice after the crystallization
can block the C+C reaction and shift
the carbon ignition to
higher densities.

\end{enumerate}

Our consideration in this paper was general.
More quantitative nuclear network simulations 
involving thermonuclear and pycnonuclear burning  
in dense stellar matter
are currently in progress.

\begin{acknowledgments}
We are grateful to 
H.~E.~DeWitt for comments and suggestions and to
M.~E.~Gusakov for providing the tables of neutrino emissivities due to
plasmon decay. This work was partially supported by The Joint Institute for
Nuclear Astrophysics (JINA) NSF PHY 0216783,
by the Russian Foundation for Basic Research
(grants 05-02-16245, 05-02-22003) and by the Federal Agency
for Science and Innovations (grant NSh 9879.2006.2).
\end{acknowledgments}

\end{document}